\documentclass[preprint]{aastex}
\usepackage{epsfig,graphicx}

\slugcomment{\sl To appear in {\it Icarus}; this preprint: Dec 02, 2003}

\def\spose#1{\hbox to 0pt{#1\hss}}
\def\gta{\mathrel{\spose{\lower 3pt\hbox{$\mathchar"218$}}
     \raise 2.0pt\hbox{$\mathchar"13E$}}}

\begin{document}

\title{ Secular Dynamics of the Three-Body Problem:\\
Application to the ${\upsilon}$ Andromedae Planetary System}

\author{T.A. Michtchenko}
\affil{Instituto de Astronomia, Geof\'{\i}sica e Ci\^encias Atmosf\'ericas, 
USP, Rua do Mat\~ao 1226, 05508-900 S\~ao Paulo, Brasil;
E-mail: tatiana@astro.iag.usp.br}
\author{R. Malhotra}
\affil{Department of Planetary Sciences, University of Arizona, 1629 East 
       University Boulevard, Tucson, AZ 85721;
E-mail: renu@lpl.arizona.edu}

\begin{abstract} 
The discovery of extra-solar planetary systems with multiple planets in
highly eccentric orbits ($\sim 0.1-0.6$), in contrast with our own Solar 
System, makes classical secular perturbation analysis very limited. In 
this paper, we use a semi-numerical approach to study the secular behavior 
of a system composed of a central star and two massive planets in co-planar 
orbits. We show that the secular dynamics of this system can be described 
using only two parameters, the ratios of the semimajor axes and the planetary 
masses. The main dynamical features of the system are presented in geometrical 
pictures that allows us to investigate a large domain of the phase space of 
this three-body problem without time-expensive numerical integrations of the 
equations of motion, and without any restriction on the magnitude of the 
planetary eccentricities.  The topology of the phase space is also investigated 
in detail by means of spectral map techniques, which allow us to detect the 
separatrix of a nonlinear secular apsidal resonance. Finally, the qualitative 
study is supplemented by direct numerical integrations. Three different regimes 
of secular motion with respect to the secular angle $\Delta\varpi$ are possible: 
they are circulation, oscillation (around $0$ and $180^\circ$), and high 
eccentricity libration in a nonlinear secular resonance. The first two regimes 
are a continuous extension of the classical linear secular perturbation
theory; the last is a new feature, hitherto unknown, in the secular
dynamics of the three-body problem. 

We apply the analysis to the case of the two outer planets in the $\upsilon$ 
Andromedae system, and obtain its periodic and ordinary orbits, the general 
structure of its secular phase space, and the boundaries of its secular 
stability; we find that this system is secularly stable over a large domain 
of eccentricities.  Applying this analysis to a wide range of planetary mass 
and semimajor axis ratios (centered about the $\upsilon$ Andromedae parameters), 
we find that apsidal oscillation dominates the secular phase space of the 
three-body coplanar system, and that the nonlinear secular resonance is
also a common feature.
\end{abstract} 



 
\section{Introduction} 

At the present time, at least ten sun-like stars in our galactic neighborhood 
are known to harbor planetary systems with two or more 
planets\footnote{http://www.obspm.fr/planets}. Two (perhaps three) of these 
include two planets locked or close to orbital mean motion resonances, and in 
these cases the orbits of the planets are significantly perturbed on orbital 
timescales by the mutual gravitational interactions of the planets. In all other 
non-resonant systems, the long term orbital variations are qualitatively and 
quantitatively described primarily by the secular effects of the gravitational 
interactions of the planets. The secular interactions can produce interesting 
orbital dynamical effects, such as alignment (or anti-alignment) of the lines of 
apsides and large amplitude variation of the eccentricities.  The planetary system 
of $\upsilon$ Andromedae has received particular attention in this regard, owing to 
the very small difference in the arguments of periastron and the surprisingly large 
eccentricities of its two outer planets (Chiang et al. 2001, Malhotra 2002, Chiang 
and Murray 2002). An understanding of the secular dynamics may potentially help
identify the origin of the large orbital eccentricities of some extra-solar planets 
(Malhotra 2002).

For the planets of our solar system (excluding Pluto), there exists a well-developed 
theoretical framework, based upon the classical Laplace-Lagrange perturbation theory 
(Laplace 1799; see, e.g., Murray and Dermott 1999) for the analysis of the secular 
orbital perturbations of the planets. However, in contrast with our solar system, the 
known exo-planetary systems have generally larger planet masses, smaller semimajor axes, 
and larger orbital eccentricities; these factors, particularly the larger eccentricities, 
limit the application of classical linear secular perturbation theory for these systems.

In the present paper, we analyze the secular dynamics of the three body problem 
consisting of a central star and two coplanar massive planets with large orbital 
eccentricities.  Our approach is, in essence, the nonlinear generalization --to 
arbitrary eccentricities-- of the Laplace-Lagrange secular perturbation theory.
We use a semi-numerical approach, which employs a numerical averaging of the short 
period gravitational interaction of the planets, to determine the interaction 
Hamiltonian describing the secular perturbations in the system.  We then use this 
{\it secular Hamiltonian} and a Hamilton-Jacobi approach to discern the geometry 
of the phase space and the main dynamical features of the system. We show that the 
secular phase space structure of the three body (two planet) system is determined 
by only two parameters, the ratio of the planet masses and the ratio of their 
orbital semimajor axes. We calculate the periodic orbits, their stability 
characteristics, and the domain of oscillation about the periodic orbits. We present 
our analysis over a wide range of planetary mass and semimajor axis ratios, including 
those pertinent to the $\upsilon$ Andromedae planetary system. 

In the low-to-moderate eccentricity regime, there exist two periodic orbits which 
can be identified with the aligned and anti-aligned modes of linear secular theory 
(the alignment refers to the longitudes of periapse of the two planets). Small 
amplitude oscillations about these periodic orbits are sometimes referred to as 
`secular resonance' or `apsidal resonance' in the literature. However, in the 
present work, we reserve the ``secular resonance'' terminology for a different 
type of motion, for the following reason. In the regime of moderate-to-high 
eccentricities, we find that the aligned mode bifurcates into a pair of stable and 
unstable periodic orbits, and a new, third, regime of motion becomes possible. 
To the best of our knowledge, this is a previously unknown feature of the secular 
dynamics of two-planet systems. It consists of a zero-frequency separatrix bounding 
a nonlinear resonance zone in the secular phase space at large values of the 
eccentricities. [An analogous feature in the secular dynamics of massless particles 
in the solar system has been described previously in Malhotra (1998).] We verify 
and supplement our analysis of the numerically averaged system with direct numerical 
integrations of the unaveraged system.

The paper is organized as follows. In section 2, we describe our model, the 
numerical averaging approach, and the geometrical method for locating the 
periodic orbits of the system using the global constants of motion. In section 
3, we apply this approach to obtain a picture of the secular phase space 
neighborhood of the outer two planets of $\upsilon$ Andromedae.  In section 4, 
we explore the secular phase space structure over a wider range of the parameter 
space of planetary masses and semimajor axis ratios. Finally, we provide a summary 
and discussion of our results in section 5.

\section{Secular dynamics in the three-body problem}

\subsection{The model} 
 
Consider two planets of masses $m_1$ and $m_2$ orbiting a central star 
of mass $M$. Hereafter, the index $i=1$, $2$ stands for the inner and outer 
planet, respectively. In the heliocentric reference frame, the set of canonical 
variables introduced by Poincar\'e (1897) consists of the planets' position vectors 
$\vec{r}_i$ relative to the star and their conjugate momenta 
$\vec{p}_i=m_i\,\frac{{\rm d}\vec{\rho}_i}{{\rm d}t}$, where $\vec{\rho}_i$ are the 
position vectors relative to the center of gravity of the three-body system (Yuasa 
and Hori 1979, Laskar and Robutel 1995). In these coordinates, the Hamiltonian $H$ 
of the three-body problem may be written in the form 
\begin{equation} 
H=\underbrace{\sum_{i=1}^2 (\frac{\vec{p}^{\,2}_i}{2\,m^\prime_i}- 
\frac{\mu_i\,m_i^\prime}{|\vec{r}_i|})}_{{\rm Keplerian\, part}}- 
\underbrace{\frac{G\,m_1\,m_2}{\Delta}}_{{\rm direct\, 
part}}+\underbrace{\frac{(\vec{p}_1\cdot \vec{p}_2)}{M}}_{{\rm indirect\, 
part}}, 
\label{eq:eq1} 
\end{equation} 
where $G$ is the gravitational constant, $\mu_i=G(M+m_i)$, 
$m_i^\prime=m_i\,M/(M+m_i)$ and $\Delta = |\vec{r}_1-\vec{r}_2|$. 
The first term produces Keplerian motions of the planets; the second 
and third terms produce direct and indirect perturbations among 
the planets, respectively. 
 
Associated with the Keplerian part of the Hamiltonian, a set of mass-weighted 
Poincar\'e elliptic variables is introduced as 
$$ 
\begin{array}{r@{=}lc@{=}c} 
\lambda_i & {\rm mean\,\,longitude,} & L_i  &m_i^\prime\sqrt{\mu_ia_i}{\rm ,}\\ 
-\varpi_i & {\rm longitude\,\,of\,\,perihelion,} & L_i-G_i  &L_i(1-\sqrt{1- 
e_i^2}){\rm ,}\\ 
-\Omega_i& {\rm longitude\,\,of\,node,} & G_i-H_i & L_i\sqrt{1-e_i^2}\,(1- 
\cos\,i_i){\rm ,} 
\end{array} 
$$ 
where $a_i$, $e_i$ and $i_i$ are the heliocentric canonical semi-major axes, 
the eccentricities and 
the inclinations of the planets, respectively. For the $\upsilon$ Andromedae system, 
dynamical stability considerations suggest mutual inclinations less than $20^\circ$ 
(Stepinski et al. 2000, Chiang et al 2001). Hence, it is reasonable, in first 
approximation, to set the planetary inclinations to zero. The Hamiltonian 
(\ref{eq:eq1}) can then be written, in terms of canonical elliptic variables, as 
\begin{equation} 
H=-\sum_{i=1}^2 \frac{\mu_i^2\,m_i^{\prime\,3}}{2\,L_i^2} - 
\frac{G\,m_1\,m_2}{a_2}\times R(L_i,L_i-G_i,\lambda_i,\varpi_i), 
\label{eq:eq2}
\end{equation} 
where the sum describes Keplerian motions of the planets and 
$R$ is the disturbing function. 

To study the secular behavior of the system, we perform a canonical transformation 
to the set of the new variables: 
\begin{equation}
\begin{array}{cll} 
\lambda_1                      & & L_1 \\ 
\lambda_2                      & & L_2 \\ 
\Delta\varpi=\varpi_2-\varpi_1 & &K_1=L_1-G_1 \\ 
-\varpi_2                      & &K_2=(L_1-G_1)+(L_2-G_2). 
\end{array} 
\label{eq:eq3}
\end{equation} 
Then we apply the following averaging procedure to the Hamiltonian system of 
Eq.~(\ref{eq:eq2}). The averaging is done with respect to the mean longitudes of 
the planets $\lambda_i$ and removes from the problem all the short periodic 
oscillations. This averaging is done by a numerical process through the double 
integration of the function (\ref{eq:eq2}). Because the Keplerian part does not 
depend explicitly on the angles or on $K_i$, it contributes 
only an inessential constant to the secular Hamiltonian, and can be neglected;
thus, the secular Hamiltonian is defined by 
\begin{equation}
{\overline H_{\rm sec}} = -\frac{1}{{(2\pi)^2}}\int_0^{2\pi}\int_0^{2\pi}
\frac{G\,m_1\,m_2}{a_2}
\times R\,d\lambda_1d\lambda_2.
\label{eq:eq4}
\end{equation}
In practice the numerical integration need be done over only the direct part of 
the disturbing function, because the indirect part, Eq.~(\ref{eq:eq1}), does not 
contain secular terms (Murray \& Dermott 1999). The value of the Hamiltonian and 
its derivatives are computed using Eq.~(\ref{eq:eq4}) for any given point of the 
phase space. We emphasize that, in the present work, we do not perform the 
developments of $H$ in the power series of the planetary eccentricities nor in 
Fourier series of the angular variables, such as in many previous analyses 
(e.g., Brouwer and Clemence 1961). This means that the secular motion of the 
planets is described very precisely with our method, without any restriction about 
the magnitude of their eccentricities.  The conditions for the applicability of the 
analysis are that the system is sufficiently far from a strong mean motion resonance 
and that the orbits are assumed to be coplanar.

After the elimination of the short periodic terms, the averaged Hamiltonian 
${\overline H_{\rm sec}}$ does not depend on $\lambda_1$ and $\lambda_2$; 
consequently, $L_1$ and $L_2$ (thus $a_1$ and $a_2$) are constant in time
and serve simply as parameters in ${\overline H_{\rm sec}}$.
In addition, due to the D'Alembert's rule, the $\varpi$-dependence in the
averaged disturbing function is only through $\Delta\varpi$. So, $-\varpi_2$
is a cyclic coordinate in ${\overline H_{\rm sec}}$, hence its conjugate momentum 
$K_2$ is also a constant of motion. Thus, the averaged planar three-body 
problem is separable and integrable.  Hamilton's equations for this problem are
as follows:
\begin{equation} 
\begin{array}{l@{=}c@{;}l@{=}c} 
\dot{K}_1& -\frac{\partial {\overline H_{\rm sec}}}{\partial \Delta\varpi} & \hspace{0.5cm} 
\Delta\dot{\varpi}& \frac{\partial {\overline H_{\rm sec}}}{\partial K_1},  
\end{array} 
\label{eq:eq5} 
\end{equation} 
and
\begin{equation} 
\begin{array}{l@{=}c@{;}l@{=}c} 
\dot{K}_2& 0& \hspace{0.5cm} 
\dot\varpi_2& -\frac{\partial {\overline H_{\rm sec}}}{\partial K_2}.  
\end{array} 
\label{eq:eq5_1} 
\end{equation} 
The solution for $K_2$ is trivial, $K_2=K^\ast_2=const$. This means that the 
averaged planar three-body problem has been reduced to a one-degree-of-freedom 
dynamical system (\ref{eq:eq5}), in which $K_2$ (and $L_1$ and $L_2$) are 
constant parameters. One important conclusion that follows is that {\it the 
variations of $\Delta\varpi$ and $K_1$ are described by a single frequency}.
\footnote{This result can also be derived from the classical Laplace-Lagrange 
secular theory for two planets.} The solution for $K_1$ and $\Delta\varpi$ can 
be obtained by integrating simultaneously the equations (\ref{eq:eq5}), for 
given fixed $K^\ast_2$ (and $L_1$ and $L_2$). Next, the equation (\ref{eq:eq5_1}) 
of motion for $\varpi_2$ can be integrated through a simple quadrature, using the 
fixed value of $K_2$ and the obtained functions $K_1(t)$ and $\Delta\varpi(t)$. 
Note that this reveals the second proper frequency in the planar secular system.
Thus, the general solution of the full two-degrees-of-freedom dynamical system 
described by ${\overline H_{\rm sec}}$ in Eq.\,\ref{eq:eq4}, is quasi-periodic, 
with two proper frequencies. We will see that particular solutions, when the two 
planets precess in lockstep with fixed eccentricities and aligned (or anti-aligned) 
apsidal lines, are periodic solutions of the full problem, but are fixed points 
of the reduced one-degree-of-freedom system, Eq.~(\ref{eq:eq5}).

After obtaining the solution in terms of the canonical variables, we can
use the expressions in Eq.~(\ref{eq:eq3}) to do the inverse transformation 
to the orbital elements to describe the secular time variations in the 
individual eccentricities and apsidal longitudes of the two planets. 
To represent the results of our study in familiar orbital elements, rather 
than the canonical set $(K_1,\Delta\varpi;K_2,-\varpi_2$), we will use the 
particular set of variables, $e_1\exp{i\Delta\varpi}$ and $e_2\exp{i\Delta\varpi}$.
This set of variables is preferable over the classical variables
($e_1\exp{i\varpi_1}$ and $e_2\exp{i\varpi_2}$) as the qualitative aspects
(alignment of lack thereof of the apsides) are more readily obtained with
these variables. To obtain the time variations of these, we need to solve only 
the first set of Hamilton's equations, Eq.~\ref{eq:eq5}. The possible solutions 
are either fixed points or periodic solutions, characterized by a single frequency. 
However, as noted above, solving the equation of motion for $\varpi_2$ 
(Eq.~\ref{eq:eq5_1}) by quadrature and the transformation to the orbital elements, 
allows us to return to the full two-degrees of freedom system and the classical 
variables $e_1\exp{i\varpi_1}$ and $e_2\exp{i\varpi_2}$. This full system has 
two proper frequencies. Keeping this in mind, we refer hereafter to the fixed 
points of Eq.~\ref{eq:eq5} as {\it periodic solutions} (of the full problem), 
and to the general solutions of Eq.~\ref{eq:eq5} as {\it quasi-periodic 
(or ordinary) solutions}.

The classical Laplace-Lagrange planetary theory (see Murray and Dermott 1999) 
describes well the behavior of the secular system, for small values of the planet 
eccentricities. The approach used in this work is an extension of the linear 
approximation to the domain of the moderate-to-high eccentricities. Hence it is 
expected that the main features of secular motion provided by the classical theory 
will be observed. Over the whole paper, we will compare the results obtained 
to those derived from the Lagrange-Laplace approach, in order to determine
the domains of validity of the linear approximation.  

The main features of secular motion of planetary systems are defined by such global 
quantities as total energy ($E$), total angular momentum ($TAM$), and Angular Momentum 
Deficit ($AMD$)(Poincar\'e 1897, Laskar 2000): 
$$
\begin{array}{ccl}
E & = & -\sum_{i=1}^2 \frac{\mu_i^2m_i^{\prime\,3}}{2\,L_i^2} + {\overline H_{\rm sec}}\\
TAM & = & L_1+L_2-K_2 \\
AMD & = & K_2.
\end{array}
$$ 
Due to the fact that, in the secular problem, semimajor axes are constants of motion, 
AMD and TAM are equivalent. AMD has a minimum value (zero) for circular orbits and 
increases with increasing eccentricities. The behavior of TAM is reverse of that of AMD.

In order to apply the developed model to the planets $C$ and $D$ of the $\upsilon$ 
Andromedae system, we adopt the following parameters: the masses $M=1.3M_{\rm Sun}$, 
$m_1=1.83M_{\rm Jup}$ and $m_2=3.97M_{\rm Jup}$, for the central star, inner planet 
C and outer planet D, respectively; the semi-major axes $a_1=0.83\,{\rm AU}$ and 
$a_2=2.5\,{\rm AU}$. These assume that both planetary orbits are observed edge-on, 
so $\sin i=1$ for each planet, where $i$ is the orbital inclination to the plane 
of the sky. Although the observational data do not constrain the orbital inclinations,
dynamical arguments suggest that the uncertainty in the individual planetary 
masses is less than a factor of $\sim2$ (Stepinski et al.~2000, Chiang et al.~2001). 
As we will see in the next section, of all these parameters, only two are 
important for the phase space structure: the semimajor axes ratio, $a_1/a_2$, 
and the mass ratio, $m_1/m_2$. The description of the secular behavior of this system 
is completed by the initial values of $e_1$, $e_2$ and $\Delta\varpi$. We adopt the 
following updated values (quoted in Chiang \& Murray 2002): $e_1=0.252$, $e_2=0.308$ 
at $\Delta\varpi=0$; formal uncertainties in the eccentricities are on the order of 20\%.  
However, we must be mindful that the observational uncertainties in orbital 
parameters of the known multiple-planet extra-solar systems are not well
determined (Ford 2003).

\subsection{Energy and AMD Levels on the Representative plane}

In this section we present a qualitative geometrical analysis of the Hamiltonian 
system, Eq.~(\ref{eq:eq4}). We begin this analysis by plotting the energy and AMD 
level curves in the space of initial conditions. The space of initial 
conditions of the Hamiltonian given by Eq.~(\ref{eq:eq4}) is three-dimensional, 
but the problem can be reduced to the systematic study of a representative 
2--D plane instead of the 3--D space as follows. In the integrable secular problem, 
the angular variable $\Delta\varpi$ can either circulate or oscillate (about $0$ or 
$180^\circ$). In both cases, it goes through either $0$ or $180^\circ$ for all 
initial conditions. Hence, without loss of generality, the angular variable 
$\Delta\varpi$ can initially be fixed at one of these values. The space of initial 
conditions can then be presented in the ($e_1$,$e_2$)--plane of initial eccentricities, 
where the initial value of the angle $\Delta\varpi$ is fixed at either zero or $180^\circ$. 

\begin{figure}
\epsscale{0.6}
\plotone{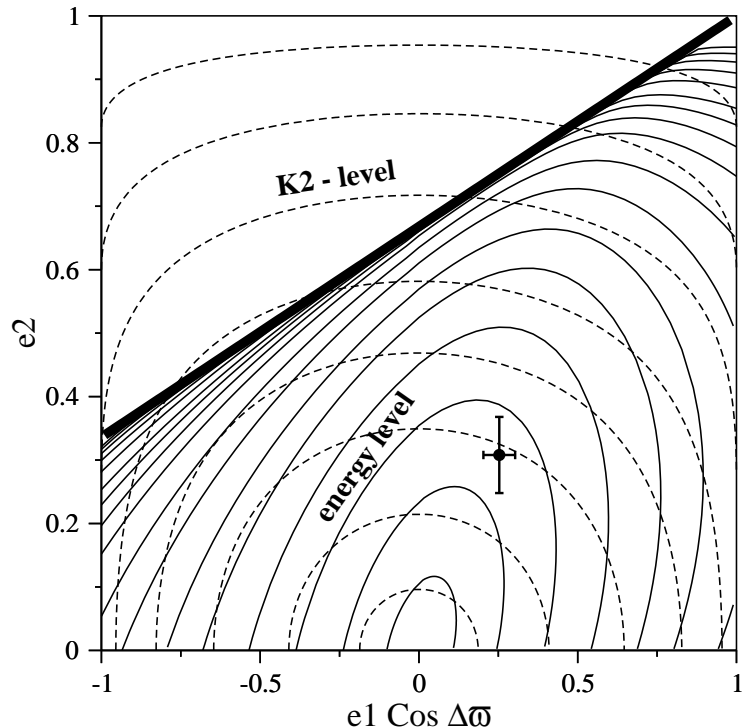}
\caption{
Energy and $K_2$ -- levels of the secular Hamiltonian given by 
Eq.~(\ref{eq:eq4}) on the ($e_1$,$e_2$) representative plane of initial 
conditions.  The energy decreases whereas $K_2$ increases with increasing 
$|e_1\cos\Delta\varpi|$ and $e_2$. The signs + or --, preceding the variable 
$e_1\cos\Delta\varpi$, indicate that the initial values of $\Delta\varpi$ 
are zero or $180^\circ$, respectively.
In constructing the level curves, we adopted the following values for the 
parameters: planet mass ratio $m_1/m_2=0.483$ and semimajor axes ratio 
$a_1/a_2=0.332$ (corresponding to those of the $\upsilon$ Andromedae planets 
C and D).
The empty region above the thick line is the domain of crossing planetary
orbits. The initial conditions of the $\upsilon$ Andromedae system in this
plane is indicated by a full circle symbol; nominal 20\% uncertainties are 
indicated for the eccentricities by error-bars.}
\label{fig1}
\end{figure}

Figure \ref{fig1} shows the levels curves of the energy and AMD 
(hereafter denoted as $K_2$--levels) on the ($e_1$,$e_2$) representative
plane. These level curves were constructed using the parameters of the  
$\upsilon$ Andromedae system of planets C and D (assuming edge-on coplanar 
orbits). The energy levels are shown by solid lines, 
whereas the $K_2$--levels are shown by dashed lines. The domain above 
the thick line at high eccentricities is the region where the planetary 
orbits cross each other; since there are no mechanisms which can protect 
the planets from close approaches between them (in contrast with the case
of mean motion resonances), this domain was excluded from our studies. 
Finally, the location of the $\upsilon$ Andromedae system 
(planets C and D) is shown by a full circle, with uncertainties of 20\%
for the planetary eccentricities.

The structure of the energy levels is independent of the planetary masses; 
it depends upon only the ratio of the semi-major axes, $a_1/a_2$. Indeed, 
as can be seen in Eq.~(\ref{eq:eq4}), the secular energy is directly 
proportional to the product of the planetary masses and the gravitational 
constant. Since the term $1/\Delta$ in the direct part of the disturbing 
function $R$ is independent of planetary masses and the indirect part does 
not contain secular terms, the secular energy can be normalized by the 
factor $\frac{Gm_1m_2}{a_2}$. Then the normalized secular energy, and 
thereby the structure of the energy levels, does not depend on the individual 
values of the semimajor axes, but depends only on the ratio, $a_1/a_2$. 

The geometry of the $K_2$--levels, however, depends on both the semimajor axes and 
the masses. From the expression for $K_2$ given by Eq.~(\ref{eq:eq3}), we can see 
that a natural normalization factor for this parameter is the constant $L_2$.  
This normalization reduces the semimajor axes dependence to only on the ratio, $a_1/a_2$.  
The mass dependence of $K_2$ goes as $\sim{m_1\over m_2}({M+m_2\over M+m_1})^{1/2}$.
Therefore, for $m_i\ll M$, in the first order mass approximation, the mass dependence 
of $K_2$ can be reduced to only on the mass ratio of the planets, $m_1/m_2$.
For example, in the $\upsilon$ Andromedae system, the planetary masses are of order 
$10^{-3}$ (in units of the stellar mass), and the first-order approximation works 
sufficiently well in this case. On the other hand, in the case of a binary stellar 
system with one planet, for example, where the mass of the companion is comparable 
to the mass of the central star, this approximation is not valid.
For the calculations presented in this paper, we do not make this approximation,
but rather adopt the values of $M, m_1,  m_2, a_1, a_2$ of the $\upsilon$ Andromedae 
system stated in the last paragraph of section 2.1.

\subsection{Geometrical pictures of the secular behavior}

In this section, we proceed with a qualitative study of the developed model 
and give a simple geometrical interpretation of the main features of secular 
motion. It should be emphasized that all information available from this study
is obtained without integration of the equations of motion, Eq.~(\ref{eq:eq5}).
The study is based on the fact that the secular energy ${\overline H_{\rm sec}}$
and AMD (i.e.~$K_2$) are both conserved along one secular path. This is reflected
on the representative plane of initial conditions in the following way: 
Independent of whether the motion is circulation or oscillation of $\Delta\varpi$,
an orbit generally passes twice through $\Delta\varpi=0$ or $180^\circ$. 
(The exceptional case of singular orbits and periodic orbits will be described below.)
This means that an individual orbit is 
represented by two points on the ($e_1$,$e_2$)--plane: one point being the 
chosen initial condition, and the other point being its counterpart. 
Due to the energy and AMD conservation, the counterpart of one initial condition 
can be found easily as an 
intersection of the corresponding energy and $K_2$--levels.

\begin{figure}
\epsscale{0.6}
\plotone{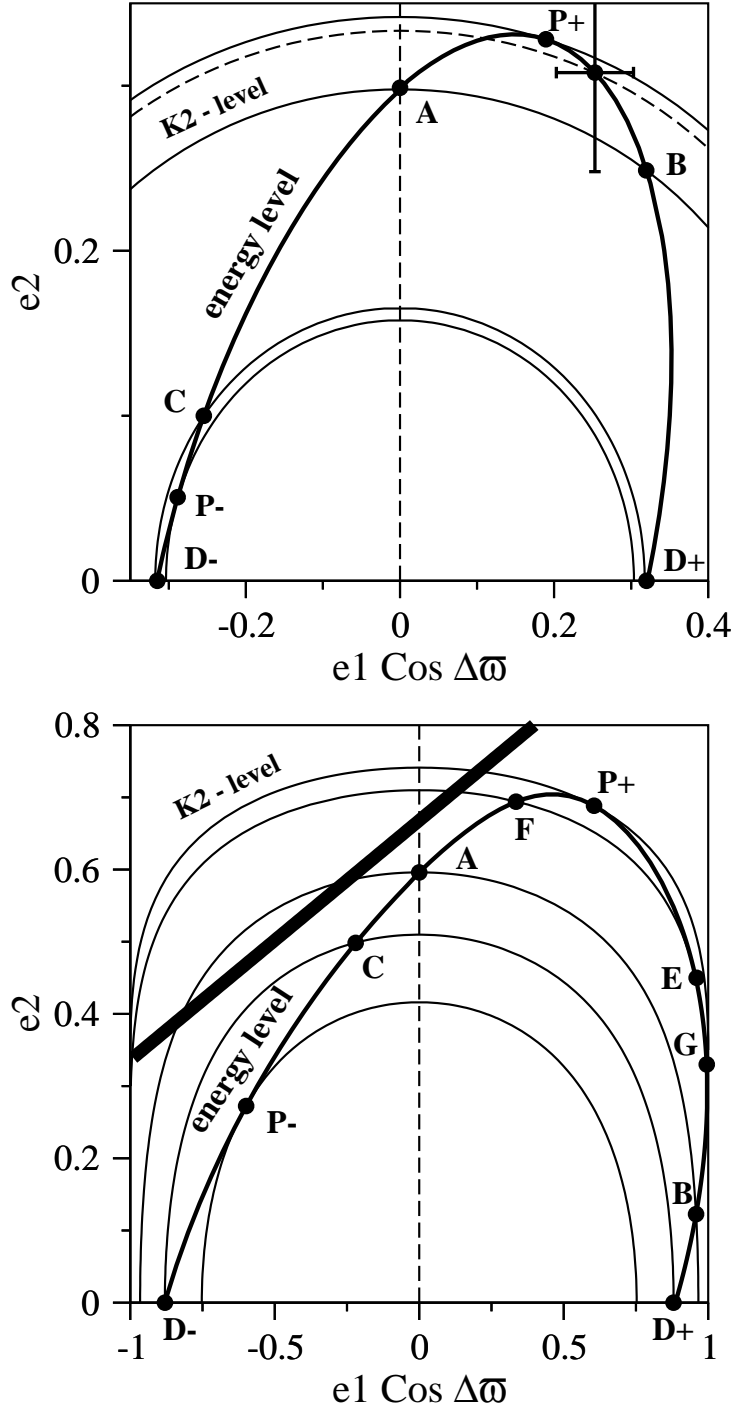}
\caption{ {\bf Top:} Energy level with ${\overline H_{\rm sec}}=-0.4 1391$
(thick curve) and four characteristic $K_2$--levels (thin curves). The location
of the current $\upsilon$ Andromedae system is indicated by a full circle 
symbol with 20\% uncertainties error-bars. The points $P^+$ and $P^-$ show 
the location of oscillation centers, around zero and $180^\circ$, respectively. The initial conditions that correspond to oscillation around zero are 
situated on the energy segment between the points {\bf $A$} and {\bf $B$}. 
Those with oscillation around $180^\circ$ between the points {\bf $C$} and 
{\bf $D^-$}. {\bf Bottom:} Energy level for ${\overline H_{\rm sec}}=-0.4295$, 
and the same four characteristic $K_2$ levels as in the top panel, plus an 
additional $K_2$--level which shows the location of a new libration center 
{\bf $E$} and its two counterparts, {\bf $F$} and {\bf $G$}. Note the 
different scales of the axes in the two panels.}
\label{fig2}
\end{figure}

Figure \ref{fig2} shows two representative ($e_1$,$e_2$)--planes of initial 
conditions which are identical to the plane shown in Fig.~\ref{fig1}, except 
that only one energy level is plotted on each of them by thick curves. In the top 
panel, the energy level corresponds to the normalized secular energy of the ``real" 
$\upsilon$ Andromedae system (${\overline H_{\rm sec}}=-0.41391$), while, in 
the bottom panel, it corresponds to a lower value ${\overline H_{\rm sec}}=-0.4295$. 
We will first concentrate our attention on the top panel. The location of the 
$\upsilon$ Andromedae system on the  ($e_1$,$e_2$)--plane is shown by a full 
circle (20\% uncertainties are indicated by error-bars).
The $K_2$--level corresponding to the system is shown by a dashed curve. This 
$K_2$--level crosses the energy level at two points: one point is the initial 
condition marked by the full circle symbol, and other point is a counterpart of the first 
point. The two points belong to the same solution of the Hamiltonian system described 
by Eq.~(\ref{eq:eq4}), and represent a quasi-periodic variation in the eccentricities 
and in $\Delta\varpi$. From the condition $K_2={\rm const}$, we can see that the 
secular oscillations of the eccentricities always occur with opposite phases: when 
the eccentricity of the outer planet $e_2$ is minimal, the eccentricity of the inner 
planet $e_1$ is maximal. Conversely, when the eccentricity $e_2$ is at a maximum, 
$e_1$ is at its minimum. 

The qualitative behavior of the angular variable $\Delta\varpi$ also can be 
obtained from the geometrical analysis of the representative 
($e_1$,$e_2$)--plane in Figure \ref{fig2}\,{\it top}. Indeed, if both points
of an orbital path are located at the positive right-hand side of 
the ($e_1$,$e_2$)--plane, the angle $\Delta\varpi$ oscillates around zero. Conversely, 
if both initial conditions are located at the negative left-hand 
side of the plane, $\Delta\varpi$ oscillates around $180^\circ$. Finally, when two 
initial conditions are at the opposite half--planes, the corresponding orbits 
of both planet are circulating with respect of the angle $\Delta\varpi$. The 
eccentricity of the more massive outer planet $e_2$ is minimal (consequently, 
$e_1$ is maximal), when $\Delta\varpi=0$, and is maximal ($e_1$ is minimal), 
when $\Delta\varpi=180^\circ$. In this way, from the geometrical picture presented in 
Figure \ref{fig2}\,{\it top}, we can calculate the ranges of the eccentricity 
variation for both planets of the system under study.

The analysis of the geometry of the energy and $K_2$--levels on the 
($e_1$,$e_2$)--plane described above is applied for any generic initial condition. 
However, for a given energy level, there are four characteristic $K_2$--levels which 
define the phase space structure and the qualitative character of the secular solutions.
These are shown in Figure \ref{fig2}\,{\it top} by continuous 
curves. The topmost $K_2$--level is tangent to the energy level at the point marked 
as $P^+$. The lowermost $K_2$--level has a minimum possible value and is tangent to 
the energy level at the point marked as $P^-$. These two $K_2$ levels represent 
extreme values of $K_2$ for the given energy level. The orbital solutions associated with 
these two extrema correspond to periodic solutions of the secular Hamiltonian system.
These periodic orbits are characterized by constant values of the eccentricity 
and of $\Delta\varpi$.

The other two characteristic $K_2$--levels contain the initial conditions
$e_1=0$ and $e_2=0$. One of these passes through the 
points {\bf $A$} and {\bf $B$} in Fig.~\ref{fig2}\,{\it top} and other one passes 
through the points {\bf $C$}, {\bf $D^+$} and {\bf $D^-$}. Note that, at $e_2=0$ the 
angular variable $\Delta\varpi$ is undefined. The points {\bf $D^+$} and 
{\bf $D^-$} both represent the initial condition $e_2=0$ on the ($e_1$,$e_2$) 
representative plane.

All initial conditions along a given energy level on the representative plane 
can be classified according to the motion of the corresponding angle $\Delta\varpi$. The 
periodic solution at point $P^+$ is characterized by zero amplitude oscillation.
The orbits which start on the segment {\bf $AB$} of the energy level are characterized 
by oscillation of $\Delta\varpi$ around zero. The center of oscillation is the point $P^+$.
The range of the eccentricity variation is given by the projections of the initial 
conditions on the corresponding axes. The periodic solution with zero amplitude oscillation
around $180^\circ$ is at $P^-$. The orbits which start on the interval between {\bf $CD^-$} 
are oscillating around $180^\circ$. Once again, the planetary 
eccentricities vary in the range given by the projections of initial conditions on the 
corresponding axes. Finally, initial conditions in the intervals {\bf $CA$} and 
{\bf $BD^+$} correspond to circulating orbits with respect to $\Delta\varpi$.

Hereafter we designate the solutions near $P^+$ as Mode I and those near $P^-$ 
as Mode II.  Near Mode I, the planetary apsidal lines are aligned, the 
secular angle $\Delta\varpi$ oscillates about $0$ and the planet eccentricities 
undergo small oscillations about the values of the Mode I periodic solution. 
Near Mode II, the apsidal lines are anti-aligned, $\Delta\varpi$ oscillates about 
$180^\circ$ and the planet eccentricities oscillate about the values of the Mode II 
periodic solution. Thus, the two modes describe two opposite stable ways of the 
planetary system to be aligned. Between the domains of Mode I and Mode II, there 
is a region of the phase space in which the motion of the angle $\Delta\varpi$ 
is a circulation. The secular oscillations of the planet eccentricities always occur 
with opposite phases. The eccentricities undergo secular variations in such way 
that,  when $\Delta\varpi=0$, the eccentricity of the outer planet is minimal 
and one of the inner planet is maximal. Conversely, when $\Delta\varpi=180^\circ$, 
the eccentricity of the outer planet D is maximal and of the inner planet is minimal.

The bottom panel  in Fig.~\ref{fig2} presents by the bold line the level curve 
of a lower 
secular energy. In this case, the domain covers higher eccentricities, and even
eccentricities close to 1. All points $P^+$, $P^-$, {\bf $A$}, {\bf $B$}, {\bf $C$}, 
{\bf $D^+$} and {\bf $D^-$} have the same interpretation as discussed above for the 
top panel. In addition, this figure exhibits a new feature which is the existence of a 
fifth characteristic $K_2$--level shown by points {\bf $E$}, {\bf $F$}, and {\bf $G$}. 
The new $K_2$--level is tangent to the energy level at point {\bf $E$}, and crosses 
it at points {\bf $F$} and {\bf $G$}. This situation means that there is one 
orbit which crosses the $e_1\cos\Delta\varpi$ axis at three different points. 
To understand the behavior in this complicated case, we will need to integrate the 
equations of motion, Eq.~(\ref{eq:eq5}); the results of the integrations are presented in the
next section.

\section{Secular Dynamics of the $\upsilon$ Andromedae system: Periodic and Ordinary 
Solutions; Nonlinear Secular Resonance} 

\begin{figure}
\epsscale{0.6}
\plotone{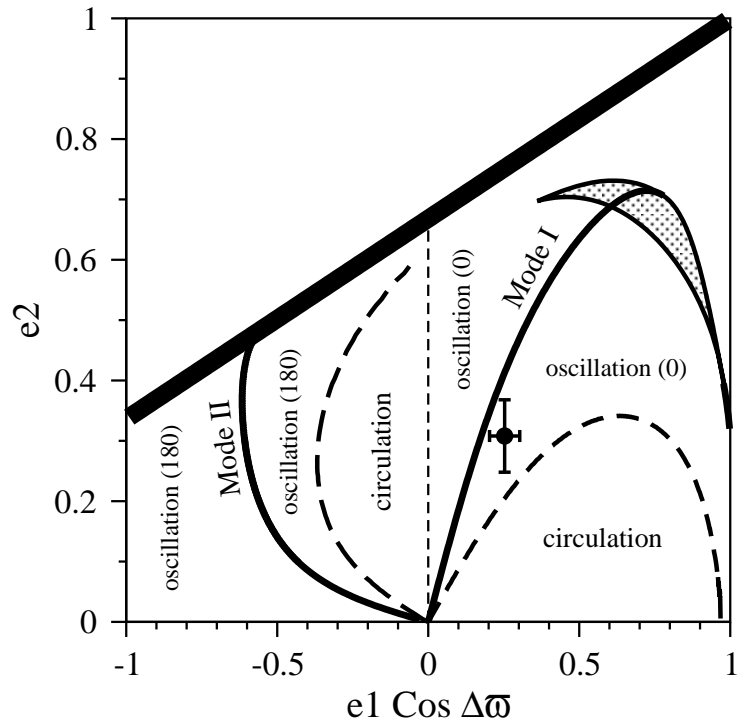}
\caption{ Periodic orbits and domains of oscillation and circulation in the
($e_1$,$e_2$)--plane of initial conditions, obtained for the parameters of the
$\upsilon$ Andromedae system. The continuous thin curves plot the location of 
the stable periodic orbits: Mode I in the positive half--plane, Mode II in the 
negative half--plane.  The domains of oscillation of $\Delta\varpi$ about $0$ 
(Mode I) and about $180^\circ$ (Mode II) are bounded by the dashed lines; the 
regions of circulation of $\Delta\varpi$ are also labeled. The shaded region 
at upper right is a family of unstable periodic orbits. The location of the 
current $\upsilon$ Andromedae system (planets C and D) is shown by the full 
circle symbol.}
\label{fig3}
\end{figure}

In this section, the insight gained by examining the geometrical structure of the 
levels on the representative plane is presented in quantitative form for the 
system of $\upsilon$ Andromedae planets C and D. Figure \ref{fig3} shows the 
locations of periodic orbits and the various domains of oscillation or circulation
of $\Delta\varpi$ in the ($e_1$,$e_2$)--plane of initial conditions, constructed for 
the parameters of this system. The detailed description and interpretation of this 
figure is as follows.

The upper boundary of the domain where the planetary orbits do not cross each other is 
shown by the thick line on the ($e_1$,$e_2$)--plane in Fig.~\ref{fig3}. This curve 
is obtained from the simple geometrical condition of two crossing orbits, 
such as: $e_2=1-a_1(1-e_1)/a_2$ and $e_2=a_1(1+e_1)/a_2-1$, 
for $\Delta\varpi=0$, and $e_2=1-a_1(1+e_1)/a_2$, 
for $\Delta\varpi=180^\circ$. Two families of stable periodic orbits are shown by the
the thin continuous curves: one is located in the positive half--plane ($\Delta\varpi=0$), 
and other in the negative half--plane ($\Delta\varpi=180^\circ$). The two half--planes are 
separated by the dashed vertical line representing the initial condition $e_1=0$.

The periodic solutions of the Hamiltonian ${\overline H_{\rm sec}}$ given by 
Eq.~(\ref{eq:eq4}) are defined by conditions:
\begin{equation} 
\begin{array}{lcl} 
\dot{K}_1=-\frac{\partial {\overline H_{\rm sec}}}{\partial \Delta\varpi}=0{\rm ,}& 
&\Delta\dot{\varpi}=\frac{\partial {\overline H_{\rm sec}}}{\partial K_1}=0{\rm .}\\                
\end{array} 
\label{eq:eq6} 
\end{equation} 
The Hamiltonian ${\overline H_{\rm sec}}$ is an even function of $\Delta\varpi$.
Hence, the first of these equations provides immediately the trivial solution 
$\Delta\varpi= 0\,\,({\rm mod}\,\,\pi)$, for non-vanishing $K_1$ and 
$K_1\neq K_2$. The second equation from Eq.~(\ref{eq:eq6}) is solved numerically for
the two cases, $\Delta\varpi=0$ and $\Delta\varpi=\pi$, and the condition $K_2={\rm const}$. 
The stability of the obtained periodic solutions is defined by the behavior of the Hessian 
matrix of ${\overline H_{\rm sec}}$ evaluated at the periodic orbit. 

Recall that in linear secular theory, the general solutions are a linear superposition of 
two linear eigenmodes (e.g., Murray \& Dermott, 1999). 
Our semi-numerical analysis yields the nonlinear generalization of the linear secular theory. 
In the positive half--plane of initial conditions in Fig.~\ref{fig3}, the domain of Mode I, 
i.e. oscillation around zero, is defined by all initial conditions between the vertical line 
$e_1=0$ and the dashed curve. The domain below the dashed curve is of circulating orbits. 
The domain of Mode II is in the negative half--plane: the initial conditions for orbits 
oscillating around $180^\circ$ lie between the horizontal axis $e_2=0$ and the dashed curve. 
The domain to the right of the dashed curve and up to $e_1=0$ is of circulating orbits. 
In the limit of linear secular theory, the two modes would be represented in Fig.~\ref{fig3} by 
straight lines tangent at the origin to the continuous curves that are the nonlinear Modes 
I and II. It is clear that the departures from the linear secular 
theory become significant for $e_1\gta 0.3$. It is interesting to note that the nonlinearity 
is not symmetric for the two modes; it is especially severe for Mode II even at relatively
small values of the eccentricity, $e_2\gta 0.1$.

Finally, the nonlinear analysis of the secular Hamiltonian yields a new feature not 
previously known, namely unstable periodic orbits at moderate-to-high eccentricities.  
In Fig.~\ref{fig3}, these correspond to the boundaries of the shaded region near the upper 
right corner.  These orbits were previously identified in Fig.~\ref{fig2} by the points 
{\bf $E$}, {\bf $F$}, and {\bf $G$}. Their interpretation requires additional information, 
which we obtain by integration of the equations of motion Eq.~(\ref{eq:eq5}). We describe 
this in more detail below, where we show that the shaded region in the ($e_1$,$e_2$)--plane 
represents a zone of nonlinear secular resonance, consisting of librating orbits bounded by 
an infinite-period separatrix. 

The location of the current $\upsilon$ Andromedae system (planets $C$ and $D$) in the 
$e_1,e_2$ plane of initial conditions is shown by the full circle symbol in Fig.~\ref{fig3}.
For the presently known best-fit planetary parameters, this system is within the Mode I 
domain with $\Delta\varpi$ oscillating about zero, as has been noted in previous studies 
(Chiang et al.~2001, Malhotra 2002, Chiang \& Murray 2002).  

\begin{figure}
\epsscale{0.6}
\plotone{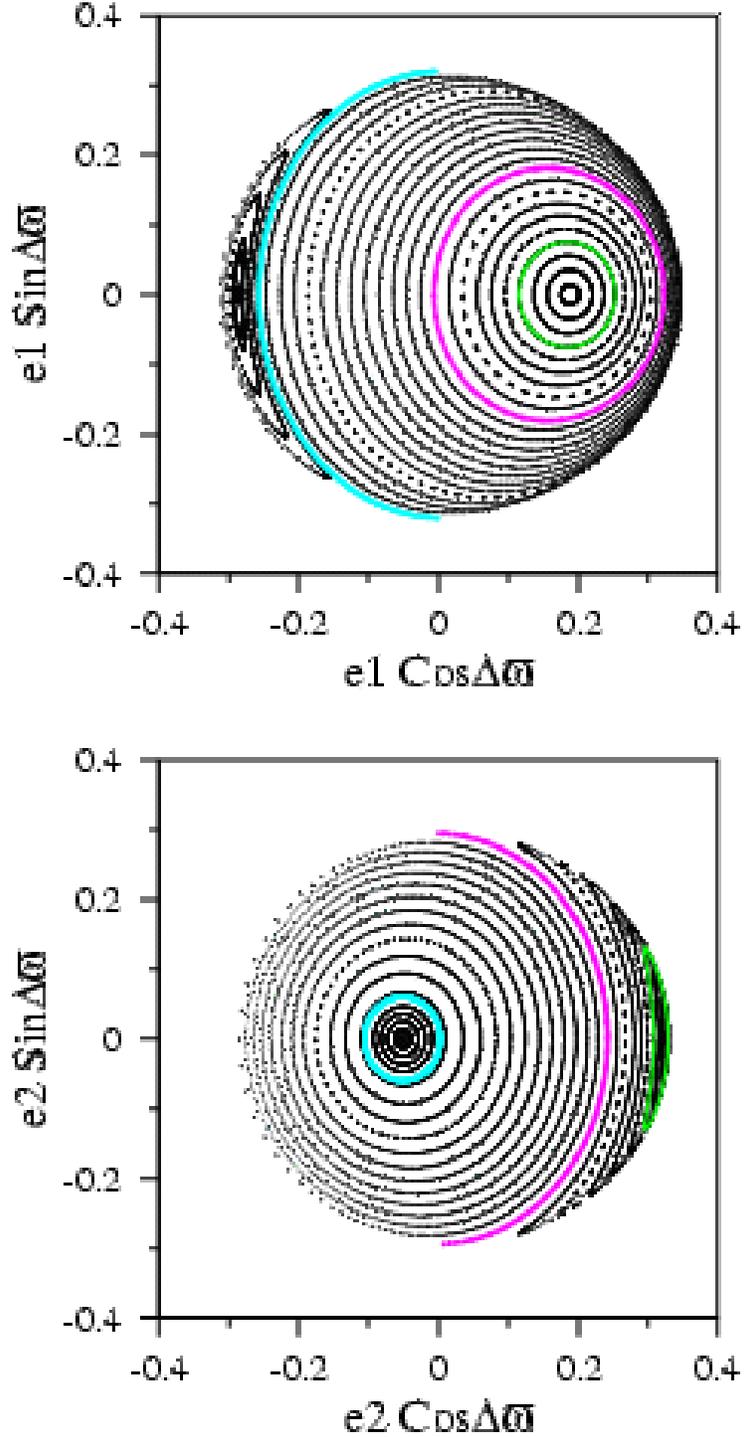}
\caption{ Secular phase space of the $\upsilon$ Andromedae planets C and D,
at ${\overline H_{\rm sec}}=-0.41391$. Locations of the Mode I and Mode II are 
given by the fixed points on the top and bottom panels. The red and blue curves 
represent orbits passing through the $e_1=0$ and $e_2=0$ singularities of 
Eq.~(\ref{eq:eq5}).  The secular variations of the $\upsilon$ Andromedae 
planets C and D for the present best-fit initial conditions are shown by 
green curves (see text).  }
\label{fig4}
\end{figure}

Figure \ref{fig4} displays the solutions of the equations of secular motion, 
Eq.~(\ref{eq:eq5}), in the two eccentricity planes, 
$(e_i\cos\Delta\varpi,e_i\sin\Delta\varpi)$. The top panel shows the secular 
variations of the inner planet and the bottom panel shows the secular variations of 
the outer planet.  These solutions were obtained by numerical integration of a large 
set of initial conditions along the energy level corresponding to the $\upsilon$ 
Andromedae system (see top panel in Fig.~\ref{fig2}). In each plane we see two fixed 
points, corresponding to the Mode I and Mode II periodic solutions. The $e_1$ plane 
is dominated by Mode I, whose fixed point lies in the central region; the Mode II fixed 
point lies near the left-hand boundary of the energy manifold. Conversely, in the 
$e_2$ plane, the Mode II periodic solution lies in the central region, while the Mode 
I periodic solution lies near the right-hand boundary of the energy manifold.

The smooth curves surrounding each of the two fixed points are periodic solutions
of the reduced one-degree-of-freedom system, but quasi-periodic solutions of the full
two-degrees-of-freedom secular system; as noted previously, we refer to these as
simply quasi-periodic solutions. Note that even though the motion of the angle $\Delta\varpi$ may be 
either oscillation (about $0$ or $180^\circ$) or circulation, there is no 
separatrix associated with an unstable infinite-period solution. To better 
understand this feature, we plot by red and blue curves two particular solutions 
in Fig.~\ref{fig4}. These solutions are associated with the singularities in 
Eqs.~(\ref{eq:eq5}), which occur at $K_1=0$ and $K_1=K_2$ or, in the 
eccentricity notation, at $e_1=0$ and $e_2=0$. (As is well known, these are not 
physical singularities, but merely due to the choice of variables.)
The solution shown by the red 
curve was obtained with initial condition $e_1 \approx 0$ and is seen as a 
smooth curve passing through the origin in the $e_1$ plane (upper panel). 
In comparison, the solution obtained with initial condition $e_2 \approx 0$ 
and shown by the blue curve, appears as the `false' separatrix in the $e_1$ 
plane, separating the domains of different modes of motion. An analogous 
situation is seen in the $e_2$ plane (lower panel), where the blue curve
is a smooth curve passing through the origin and the red curve separates 
the domains of two modes of motion. We emphasize that the boundary separating 
the oscillation of $\Delta\varpi$ about either 0 or $180^\circ$ is not a 
zero-frequency separatrix, but simply the quasiperiodic solutions that pass 
through the singular initial values, $e_1=0$ or $e_2=0$. The separatrix-like 
curves in Fig.~\ref{fig4} are owed to the effects of projecting the canonical
solutions on to a plane of non-canonical variables;
these are better visualized over a sphere (for more details, see 
Pauwels 1983).

Finally, the solution corresponding to the current best-fit initial conditions of 
the $\upsilon$ Andromedae planets $C$ and $D$ is shown by the green trajectories 
in Fig.~\ref{fig4}.  This solution oscillates about the Mode I periodic solution;
the oscillation amplitude of $\Delta\varpi$ is 25$^\circ$ about 0. Note, however, 
that the oscillation amplitude is quite sensitive to the initial conditions. 
This is most readily seen in the structure near the green curves in Fig.~\ref{fig4}: 
small changes in initial conditions lead to large changes in the oscillation 
amplitude. A complete analysis of the observational uncertainties in the parameters 
of this system is beyond the scope of the present work.  Suffice to note that  
these uncertainties (see Fig.~\ref{fig3}) do not exclude significantly larger 
or significantly smaller amplitudes of oscillation of $\Delta\varpi$. For example, 
assuming all other parameters are fixed, the 20\% nominal uncertainty in the 
eccentricities implies that the $\Delta\varpi$ oscillation amplitude may be as 
large as $70^\circ$ or as small as $9^\circ$. As we discuss in section 5, this 
uncertainty is problematic for theories of the origin of the large eccentricities.

\begin{figure}[!t]
\epsscale{0.6}
\plotone{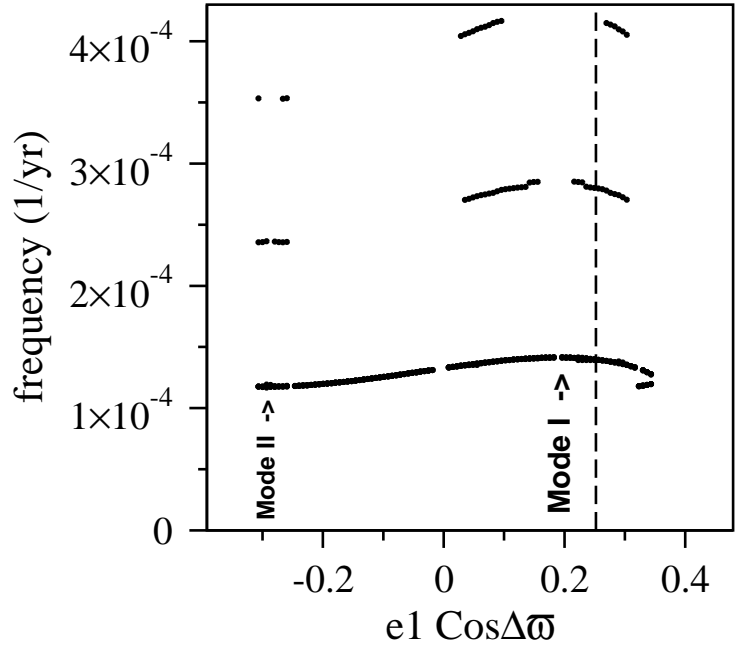}
\caption{
Spectral map corresponding to the solutions shown on the planet C panel in 
Figure \ref{fig4}. The abscissas ($e_1\cos\Delta\varpi$) in the spectral map 
are the initial conditions at $\Delta\varpi=0$ and $\Delta\varpi=180^\circ$. 
The dashed line indicates the initial value of $e_1\cos\Delta\varpi$ of the 
${\upsilon}$ Andromedae system.  }
\label{fig5}
\end{figure}

Figure \ref{fig5} shows the spectral map obtained from the Fourier analysis 
of solutions calculated with initial conditions on the horizontal axis on the 
planet C panel in Fig.~\ref{fig4}\,{\it top}. The map shows the frequency of 
all peaks present in the Fourier spectra of the eccentricities $e_1$ and $e_2$ 
that are clearly above noise level (arbitrarily defined as 20 per cent of the 
largest spectral peak). Besides the discussion above about the nature of the 
solutions separating the oscillation zones, one may note the continuous evolution 
of the frequencies on the corresponding spectral map. The map shows only one 
frequency and its higher harmonics that is characteristic of one-degree of 
freedom dynamical systems. The continuous smooth evolution of frequency through 
the domain of initial conditions does not exhibit the typical signature of a 
separatrix associated with an unstable periodic orbit. The range of the variation 
of the period of $\Delta\varpi$ is between 6,500 and 10,000 years: the shortest 
period occurs when the system is near Mode I and the longest period occurs 
near Mode II.  The initial co-ordinate of the ${\upsilon}$ Andromedae system 
(planets C and D) is indicated by the dashed vertical line; the oscillation period 
of $\Delta\varpi$ in this case is $\sim 7,100$ years.

\begin{figure}
\epsscale{0.5}
\plotone{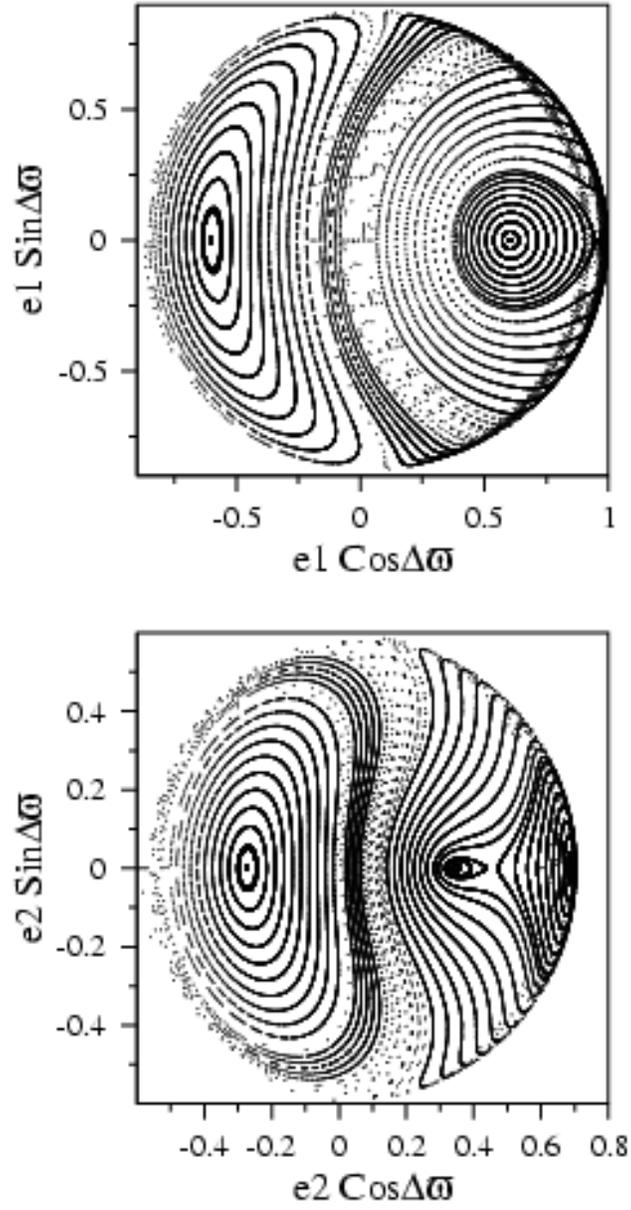}
\caption{ Same as in Figure \ref{fig4}, but for 
${\overline H_{\rm sec}}=-0.4295$.
The erratic scatter of the points at eccentricities close to 1 is due to the
loss of numerical accuracy.}
\label{fig6}
\end{figure}

\begin{figure}[!t]
\plotone{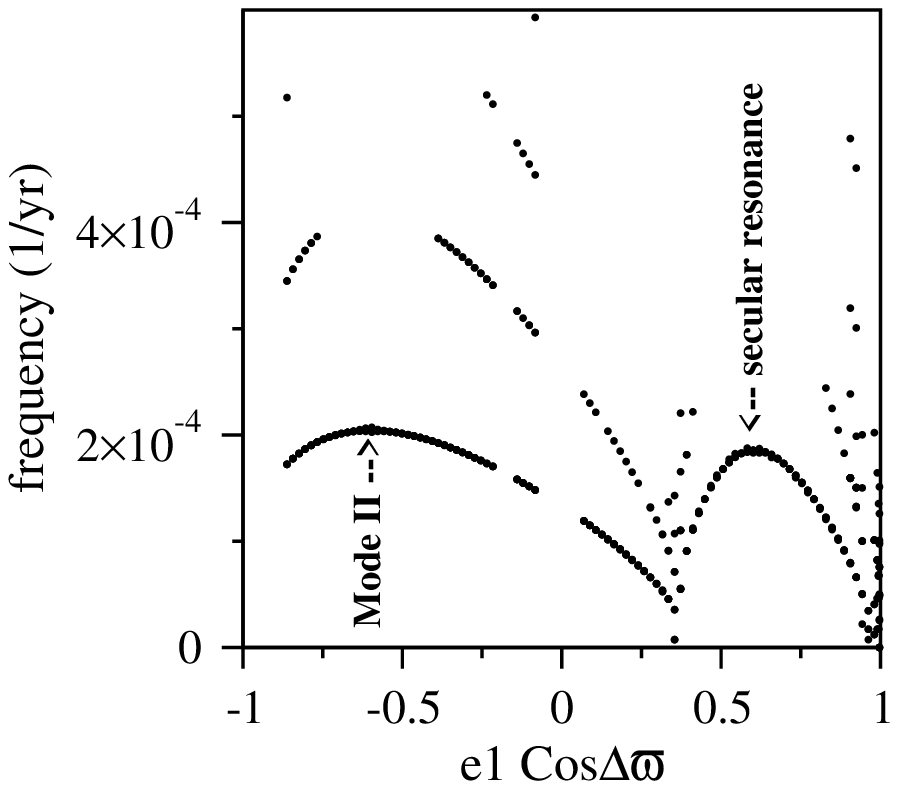}
\caption{Same as in Figure \ref{fig5}, but for 
${\overline H_{\rm sec}}=-0.4295$.}
\label{fig7}
\end{figure}

Figures \ref{fig6} and \ref{fig7} are analogous to Figures \ref{fig4} and \ref{fig5},
respectively, except they were obtained for the lower value of the normalized 
secular energy, ${\overline H_{\rm sec}}=-0.4295$. For this lower energy, we find 
that a qualitatively new feature appears in the phase space. In Fig.~\ref{fig6}, we 
see the formation of a new libration center and one saddle-point in the domain of 
high eccentricities, where the effect of nonlinearity of the secular perturbations 
is strong. This feature can be associated with the advent of a {\it nonlinear secular 
resonance} zone, which is bounded by a zero-frequency separatrix and which occupies 
the domain about the Mode I fixed point at large eccentricities. 
This feature can be identified in the spectral map shown in Fig.~\ref{fig7},
where the $\Delta\varpi$--frequency tends to $0$ at $e_1\cos\Delta\varpi=0.36$. 

\section{Dependence on the Planetary Masses and Semimajor Axes Ratios}

In the previous sections, we have analyzed the secular behavior for one set of values 
the planetary mass and semimajor axis ratios: $a_1/a_2=0.332$ and $m_1/m_2=0.461$, 
values which are pertinent to the current ${\upsilon}$ Andromedae planetary system. 
In this section, we will vary these two parameters and see the effect on the secular 
phase space structure. To see this dependence, we will plot the locations of periodic 
orbits and the domains of circulation or oscillation of $\Delta\varpi$ in the 
($e_1$,$e_2$) representative plane of initial conditions. We will first consider 
the case when the outer planet is equal or more massive than inner one, that is 
$m_1/m_2\leq 1$. Fixing the inner mass at $m_1=1.83M_{\rm Jup}$, we take three 
values of the outer mass: $m_1/m_2=1/1$, $m_1/m_2=1/2$ and $m_1/m_2=1/4$. Also, 
fixing the inner semimajor axis at $a_1=0.83\,{\rm AU}$, we take three values of 
the outer semi-major axis: $a_2=1.5\,{\rm AU}$, $a_2=2.5\,{\rm AU}$ and 
$a_2=5\,{\rm AU}$. 

\begin{figure}[!t]
\epsscale{0.6}
\plotone{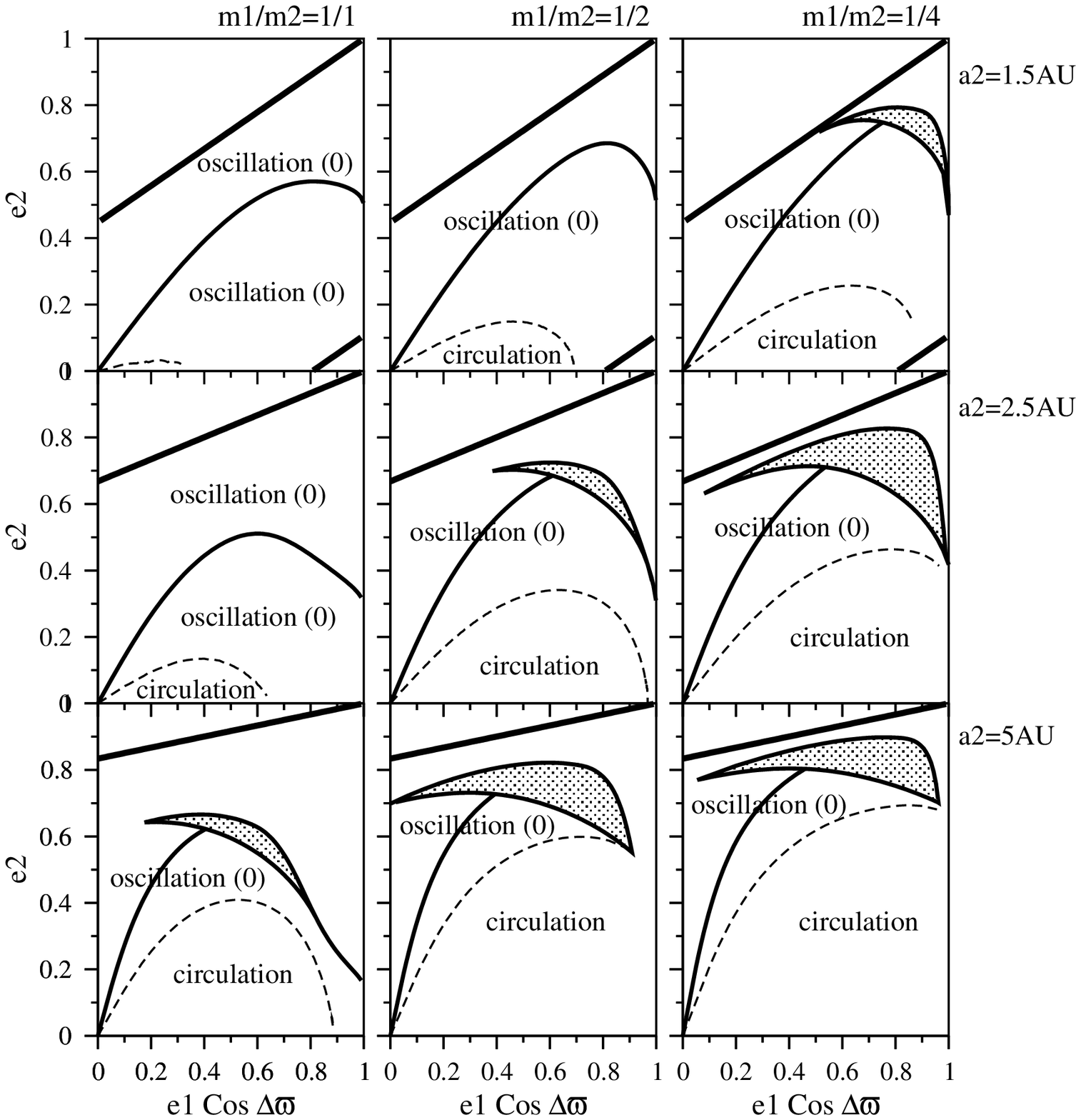}
\caption{ Periodic orbits and domains of oscillation and circulation of
$\Delta\varpi$ in the representative ($e_1$,$e_2$)--planes, for a range of 
planetary masses and semimajor axes. {\bf Rows:} a fixed value of the inner 
semi-major axis $a_1=0.83\,{\rm AU}$ and three values of the outer semi-major 
axis: $a_2=1.5\,{\rm AU}$\,(top), $a_2=2.5\,{\rm AU}$\,(middle) and
$a_2=5\,{\rm AU}$\,(bottom); {\bf Columns:} a fixed value of the inner mass
$m_1=1.83M_{\rm Jup}$ and three values of the outer mass, such as:
$m_1/m_2=1$\,(left), $m_1/m_2=1/2$\,(middle) and $m_1/m_2=1/4$\,(right). 
The shaded regions are domains of the nonlinear secular resonance.}
\label{fig8}
\end{figure}

The results are shown in Fig.~\ref{fig8}, where the rows present the variation 
with mass parameter, for a fixed value of $a_1/a_2$, while the columns present the 
variation with axis parameter, for a fixed value of $m_1/m_2$. Since the most 
significant changes occur 
at $\Delta\varpi=0$, only positive half--planes are presented in Fig.~\ref{fig8}. 
In each plane, the boundaries of the domain of non-crossing planetary orbits are 
shown by thick inclined lines. The families of stable periodic orbits are shown 
by continuous black curves, whereas the boundaries of domains of 
$\Delta\varpi$--oscillation are shown by dashed curves. Finally, the domains of the
secular resonance are shown as shaded regions on the ($e_1$,$e_2$)--planes.
Note that the parameters (planet mass and semimajor axis ratios) used in 
construction of the mid-row and mid-column are very close to those of the current 
${\upsilon}$ Andromedae system shown in Fig.~\ref{fig3}.

Figure \ref{fig8} gives a global view of all possible regimes of motion with 
respect to the secular angle $\Delta\varpi$. When the planetary masses are 
equal and the mutual distances are small, almost the whole phase space of the 
system is oscillations of the angle $\Delta\varpi$ around 0. In this 
case, as can be easily inferred from the geometrical pictures in Fig.\,2, 
the oscillations of $\Delta\varpi$ around $180^o$ will cover almost the whole 
negative half--plane of initial conditions (not shown here).
The circulation occurs only in the domains of small eccentricities. With 
increasing mutual distance of the planets, the regions of circulation are 
also increasing. The same effect is observed when the mass of the outer planet 
increases. The secular resonance seems to be a common feature of the planetary 
system, particularly, when the masses and axes ratios are not close to 1.
Its location on the ($e_1$,$e_2$) representative plane of initial conditions is 
always in the regions of the very-high eccentricities of the more massive outer
planet. The orbit of the inner planet in this case can be nearly circular.

\begin{figure}[t]
\epsscale{0.5}
\plotone{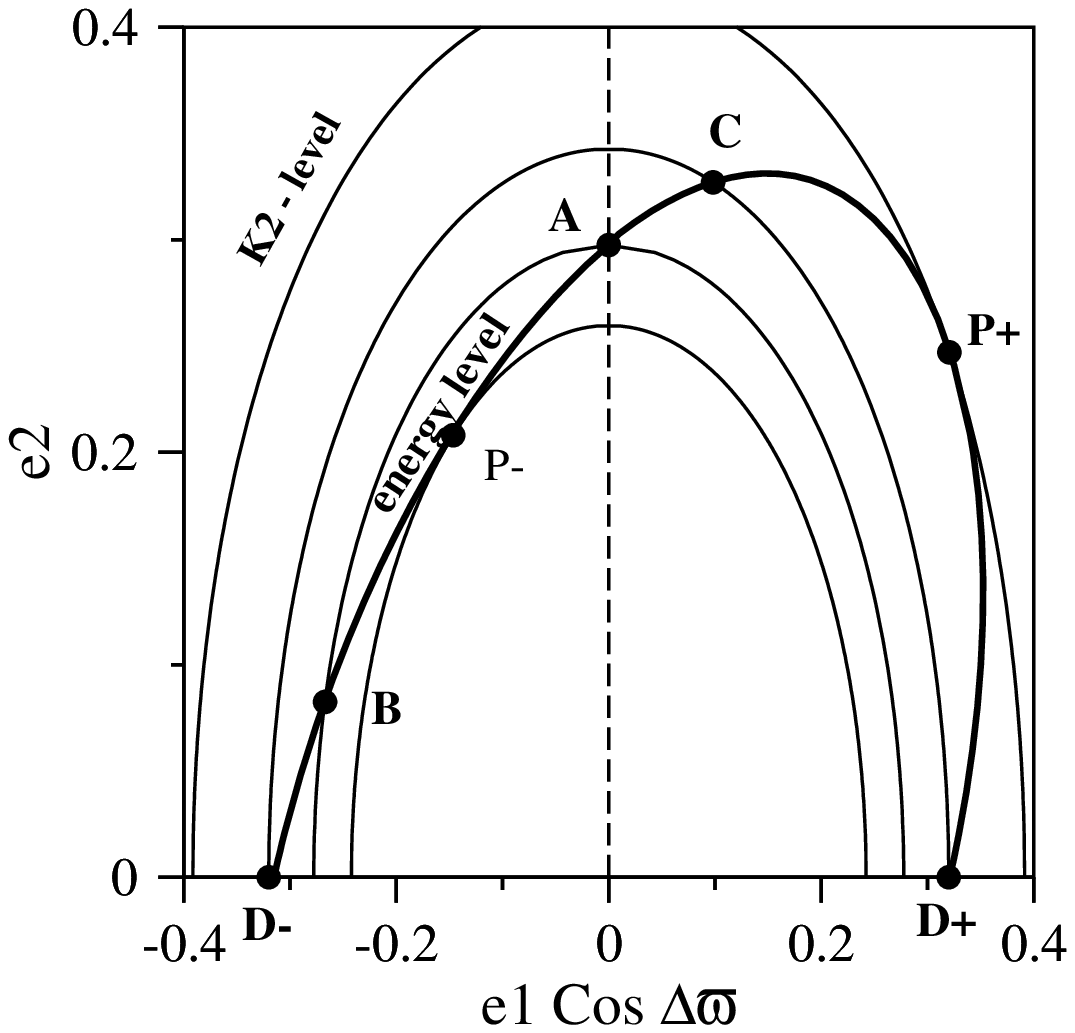}
\caption{Same as in Figure \ref{fig2}, but calculated with 
$m_1=3.97M_{\rm Jup}$ and $m_2=1.83M_{\rm Jup}$.}
\label{fig9}
\end{figure}

Now we will consider the case when mass of the inner planet is larger that the mass of 
the outer planet, that is $m_1/m_2>1$. The geometrical picture of the ($e_1$,$e_2$) 
representative plane of initial conditions corresponding to this case is shown in 
Fig.~\ref{fig9}. This plane was constructed with the same axes ratio used in 
Fig.~\ref{fig3}, but the planetary masses were changed to $m_1=3.97M_{\rm Jup}$ and 
$m_2=1.83M_{\rm Jup}$. Since the axes ratio has not been changed, the energy level 
shown by the thick curve is the same level, $H_{sec}=-0.41391$, shown in Fig.~\ref{fig2}. 
However, the structure of the $K_2$--levels is different when compared to those in 
Fig.~\ref{fig3} because of the mass dependence of the variable $K_2$. The main geometrical 
difference is the location of the points B and C on the ($e_1$,$e_2$)--plane: now the 
point B is located in the negative half--plane and the point C in the positive 
half--plane. This means that the oscillation of $\Delta\varpi$ around $0$ occurs along 
the energy segment ${\bf CD^+}$ and around $180^\circ$ along the energy segment 
${\bf AB}$. Finally, the initial conditions chosen in the intervals {\bf $CA$} and 
{\bf $BD^-$} correspond to circulating orbits with respect to $\Delta\varpi$. 

\begin{figure}[!t]
\epsscale{0.37}
\plotone{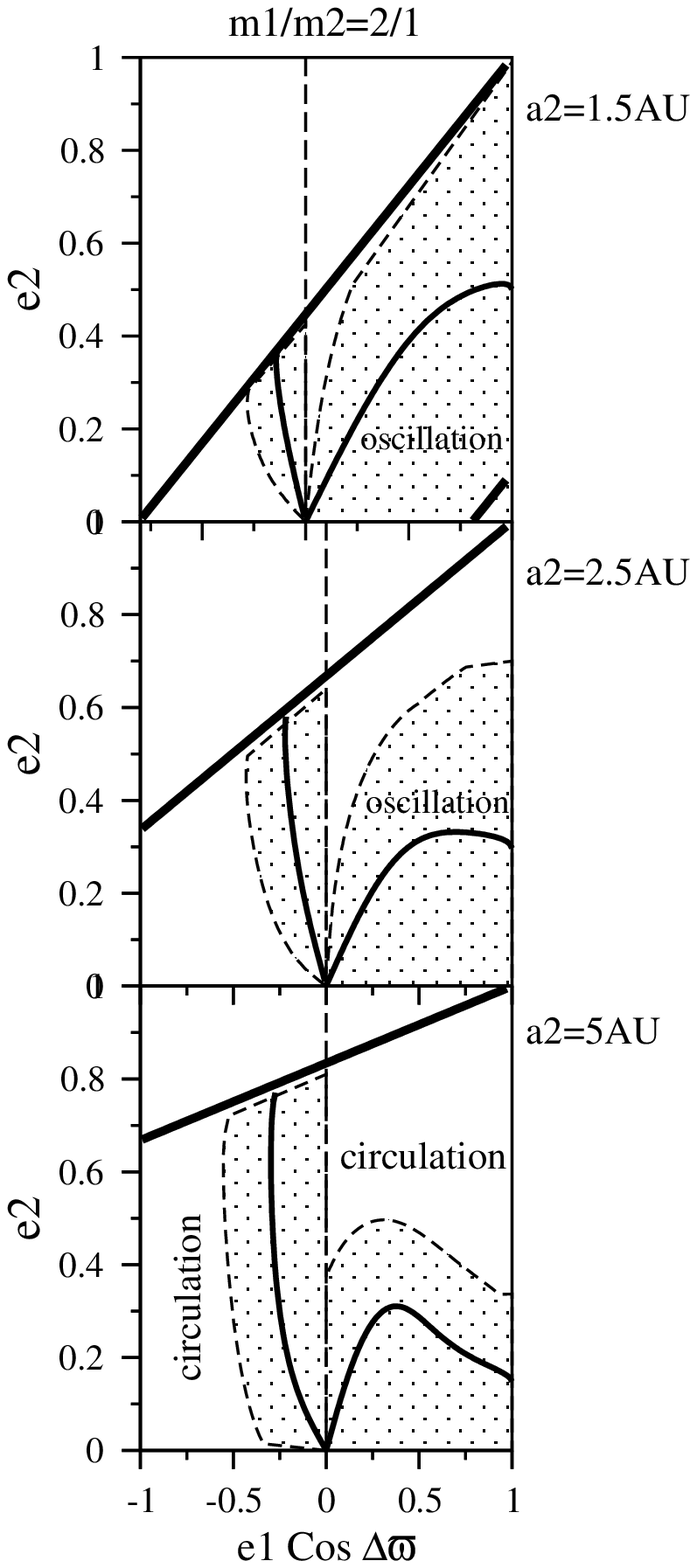}
\caption{ Same as in Figure \ref{fig8}, but for $m_1=3.97M_{\rm Jup}$ and
$m_2=1.83M_{\rm Jup}$. The shaded regions are domains of oscillations of 
$\Delta \varpi$ around $0$ or $180^o$.
}
\label{fig10}
\end{figure}


The dependence of the secular behavior on the semimajor axes ratio in the case of 
$m_1/m_2=2$ is shown in  Fig.~\ref{fig10}. This figure is similar to Fig.~\ref{fig8}: 
it was constructed using the same three values of the outer semi-major axis 
$a_2=1.5\,{\rm AU}$, $a_2=2.5\,{\rm AU}$ and $a_2=5\,{\rm AU}$, whereas the inner 
semimajor axis was fixed at $a_1=0.83\,{\rm AU}$. As in Figure \ref{fig8}, the 
boundaries of the domain of noncrossing planetary orbits are shown by the thick 
inclined lines on each panel. The families of stable periodic orbits are shown by 
the thin continuous curve, whereas the boundaries of domains of 
$\Delta\varpi$--oscillation are shown by the dashed curves. The main difference 
in the secular motion is that in this case the periodic orbits have lower values 
of $e_2$ compared to the case with $m_1/m_2=1/2$, and the nonlinear secular resonance
zone does not occur for all considered values of the axes ratio.  
Interestingly, the location of Mode I in this case is remarkably different
from that predicted by linear secular theory.

\section{Summary and Discussion}

The nonlinear secular perturbation analysis of the planar three body
(two planet) problem presented here has been motivated by the recent
discovery of extra-solar planetary systems which have large orbital
eccentricities and, in some cases, exhibit secular oscillations of
their apsidal lines.  The classical Laplace-Lagrange linear secular
perturbation theory is insufficient for the precise analysis of such 
systems. We developed a semi-analytical approach, consisting of 
a numerical averaging over the short-period perturbations in the
mutual interaction of the two planets to obtain the {\it secular} 
Hamiltonian. The Hamilton-Jacobi approach we use shows 
that the secular Hamiltonian is separable.  
(It is reducible to a one-degree-of-freedom integrable dynamical system 
for the canonical coordinate $\Delta\varpi$.)
We obtain the periodic orbits of this system, and we 
present a geometrical picture of the secular phase space of the two-planet 
system in terms of physical variables, the eccentricities and periastron 
longitudes. This phase space structure depends upon two parameters: the 
ratios of the planets' masses and their semimajor axes. The description of 
the secular variations of the system is completed by the initial conditions:
the eccentricities and the relative longitude of the periastrons. Owing to 
the symmetries in the secular Hamiltonian, the phase space structure can be 
visualized in a two dimensional representative plane of initial conditions, 
$e_2$ and $e_1\cos\Delta\varpi$, where initial $\Delta\omega$ is either 
0 or $180^\circ$. 

The locations of periodic orbits and the domains of different 
quasiperiodic motions (corresponding to the stationary solutions
and oscillations or circulations of $\Delta\varpi$)
are then easily obtained in this plane of initial conditions, for given
values of the parameters ($m_1/m_2, a_1/a_2$). As expected from the linear 
secular perturbation theory for two planets, there exist two families of stable 
periodic orbits which describe the aligned and anti-aligned orientations of the 
two planets' apsidal lines. Departures from the linear secular theory are readily 
visualized in this plane of initial conditions.  In the domain of large 
eccentricities, we find a new feature which consists of a family of unstable 
periodic orbits.  In this regime, the secular phase space contains a nonlinear 
resonance zone bounded by a zero-frequency separatrix. We emphasize that all of 
these results are obtained by means of a geometrical analysis based on the global
constants of motion, and without time-expensive numerical integrations of the 
equations of motion.

We applied this analysis to the specific case of the system of two outer 
planets, C and D, of $\upsilon$ Andromedae, assuming edge-on co-planar 
orbits. We find that the nonlinear effects are significant in this case for 
eccentricities exceeding $\sim0.3$ in the neighborhood of the aligned mode,
and $\sim0.1$ in the neighborhood of the anti-aligned mode. As noted already 
in previous studies of the dynamics of this system, the present best-fit 
initial conditions place this system in the domain of oscillations about the 
periodic orbit describing the aligned mode (Mode I, $\Delta\varpi=0$). 
Our analysis shows that the secular behavior of the $\upsilon$ Andromedae 
system is stable over 
a large domain of initial eccentricities. On the representative plane of 
initial conditions, the domain of stable motion has an upper bound at the 
collision curve defined by close approaches between two planets; below the 
collision curve, a secular instability occurs in the vicinity of the nonlinear 
secular resonance located near $e_1=0.6$ and $\Delta\varpi =0$ (Fig.~\ref{fig3}).

We have presented an analysis of the secular dynamics of the three-body 
planetary system over a wide range of the planetary mass and semimajor axis 
ratios. The systems always exhibit two main regimes of motion, characterized 
by circulation of $\Delta\varpi$ or its oscillation around 0 or $180^\circ$.  
The existence of a third regime, the nonlinear secular resonance, depends 
on the values of the physical parameters. For example, when the planetary masses 
are equal and their mutual distances are small, the nonlinear secular resonance 
does not occur; however, under these conditions, the probability of finding the 
system in the oscillation (around 0 or $180^\circ$) regime of motion is very 
close to 1. When both mass and semimajor axis ratios are far from 1, the domains 
of oscillation of $\Delta\varpi$ decrease, and the nonlinear secular resonance zone 
appears. The secular resonance feature discovered in our nonlinear
analysis is presently only a theoretical curiosity, as no real system
is known which can be associated with this dynamical state. However, the 
regime of planetary parameters where this feature is significant
is not especially extreme, and it would not be surprising if such
systems are identified in the future.  

The origin of the high planetary eccentricities of the $\upsilon$ Andromedae 
planets C and D, as well as of many other exo-planetary systems, are an
outstanding problem of high current interest.
Two basic physical mechanisms that have been proposed to account for 
the high eccentricities invoke dynamical planet-planet interactions 
(Ford et al.~2001, Marzari \& Weidenschilling 2002) or planet-disk 
interactions (Goldreich \& Sari 2003); more complex scenarios 
involving planet-disk interactions moderated by resonant
interactions between two planets have also been suggested 
(Chiang et al.~2002, Lee \& Peale 2002, Malhotra 2002, Chiang \& Murray 2002).
It has been pointed out that in some multiple-planet systems the 
occurrence of apsidal oscillation, and especially the amplitude of that 
oscillation, may serve to constrain the physical mechanisms (Malhotra 2002).
Unfortunately, the observational uncertainties of the planetary parameters of 
even the best-known exo-planetary system, $\upsilon$ Andromedae, 
are still too large to define sufficiently precisely the amplitude of 
oscillation of $\Delta\varpi$, and thus to favor one scenario over another.
We await future improvements in the accuracy of the parameters of this system 
and others to more precisely determine their dynamics and thereby constrain
their dynamical history.

The method developed here is a non-linear generalization of the
classical secular perturbation theory for the planar three body problem.
In future work we hope to extend this to non-coplanar systems, and to
apply the analysis to the secular dynamics of other planetary systems.
We note that the Hamilton-Jacobi approach used here also 
provides a powerful means to explore the effects of adiabatic evolution 
of the energy and angular momentum due to dissipative interactions
with the nascent protoplanetary disk, identify adiabatic invariants,
and thus possibly obtain additional insights into the dynamical history 
of some exo-planetary systems.

\section*{Acknowledgments}
We would like to thank Prof. Dr. S. Ferraz-Mello for critical reading through
the manuscript. TAM's work has been supported by the S\~ao Paulo State Science 
Foundation (FAPESP), as well as the Brazilian National Research Council (CNPq).
RM acknowledges research support from NASA (grants NAG5-10343 and NAG5-11661). 
The authors gratefully acknowledge the support of the Computation Center of the 
University of S\~ao Paulo (LCCA-USP) for the use of their facilities.

\vspace{-0.5cm}
\section*{\centering { \normalsize REFERENCES}}
\vspace{-0.2cm}

\begin{list}{} 
{ \setlength{\leftmargin}{0.5cm}
\setlength{\itemindent}{-0.5cm}}

\item
Brouwer, D., and G.M. Clemence 1961. Methods of Celestial Mechanics. 
Academic Press, New York and London.

\item
Chiang, E.I., S. Tabachnik, and S. Tremaine 2001. Apsidal Alignment in 
${\upsilon}$  Andromedae. {\it AJ} {\bf 122}, 1607-1615.

\item
Chiang, E.I., D. Fischer, and  E. Thommes 2002. Excitation of Orbital 
Eccentricities of Extrasolar Planets by Repeated Resonance Crossings. 
{\it ApJ} {\bf 564}, L105-L109.

\item
Chiang, E.I., and N. Murray 2002. Eccentricity Excitation and Apsidal 
Resonance Capture in the Planetary System ${\upsilon}$  Andromedae. 
{\it ApJ} {\bf 576}, 473-477.
                   
\item
Ford, E.B., 2003. Quantifying the uncertainty in the orbits of extra-solar planets,
{\it AJ}, submitted (preprint: astro-ph/0305441).
                   
\item
Ford, E.B., M. Havlikova, and F.A. Rasio 2001. Dynamical instabilities in 
extrasolar planetary systems containing two giant planets. {\it Icarus} 
{\bf 150}, 303. 

\item
Goldreich, P., and  R. Sari 2003. Eccentricity Evolution for Planets in Gaseous Disks.
 {\it ApJ} {\bf 585}, 1024-1037.

\item
Laplace, P.S. 1799. ``M\'ecanique C\'eleste". English translation by N. 
Bowditch, Chelsea Pub. Comp. Edition, N.Y., 1966.

\item
Laskar, J., and P. Robutel 1995. Stability of the planetary three-body problem:
I. Expansion of the planetary Hamiltonian. 
{\it Celest. Mech. Dynam. Astron.} {\bf 62}, 193-217.

\item
Laskar, J. 2000. On the Spacing of Planetary Systems. Physical Review Letters 
{\bf 84}, 3240-3243.

\item
Lee, M.H., and S.J. Peale 2002. Dynamics and Origin of the 2:1 Orbital 
Resonances of the GJ 876 Planets.  {\it ApJ} {\bf 567},  596-609.

\item
Malhotra, R. 1998. Orbital resonances and chaos in the Solar 
System. In {\it Solar System Formation and Evolution} ASP Conference Series 
(D.Lazzaro {\it et al.}, Eds.) {\bf 149}, 37-63.

\item
Malhotra, R. 2002. A Dynamical Mechanism for Establishing Apsidal Resonance. 
{\it ApJ}  {\bf 575}, L33-L36.

\item
Marzari, F., and S.J. Weidenschilling 2002. Eccentric Extrasolar Planets: 
The Jumping Jupiter Model. {\it Icarus} {\bf 156}, 570-579.

\item
Michtchenko, T. and S. Ferraz-Mello 2001. Modeling the 5:2 mean-motion 
resonance in the Jupiter-Saturn planetary system. {\it Icarus} {\bf 149}, 
357-374.

\item
Murray, C.D., and S.F. Dermott 1999. Solar System Dynamics. Cambridge 
University  Press, 274-279.

\item
Pauwels, T. 1983. Secular orbit-orbit resonance between two satellites
with non-zero masses. {\it Celest. Mech.} {\bf 30}, 229-247.

\item
Poincar\'e, H. 1897. Sur une forme nouvelle des \'equations 
du probl\`eme des trois corps. {\it Bull.Astron.} {\bf 14}, 53-67.

\item
Stepinski, T. F., R. Malhotra, and D.C. Black 2000. The $\upsilon$ 
Andromedae System: Models and Stability. {\it ApJ} {\bf 545}, 1044-1057.

\item
Yuasa, M., and G. Hori 1979. New approach to the planetary theory. In {\it Dynamics
of the Solar System} (R.L. Duncombe, Ed.), D.Reidel, Dordrecht, 69-73.

\end{list}
\end{document}